\documentclass[twocolumn,prl,floatfix,citeautoscript,nofootinbib,superscriptaddress]{revtex4}
\usepackage{amsbsy}
\usepackage{latexsym,epsfig,graphicx}
\usepackage{dcolumn}
\usepackage{graphicx}
\usepackage{subfigure}
\usepackage{comment}
\usepackage{color}
\usepackage{bm}
\usepackage{mathrsfs}
\usepackage{amsfonts}
\usepackage{amsmath}
\usepackage{color}
\usepackage{amssymb}
\usepackage{xspace}
\usepackage{epstopdf}
\usepackage{tabularx}
\usepackage{longtable}
\usepackage[colorlinks=true, letterpaper=true, pdfstartview=FitV, linkcolor=blue, citecolor=blue, urlcolor=blue]{hyperref}
\usepackage[normalem]{ulem}

\pdfoutput=1

\begin{document}

\author{Junpeng Hou}
\affiliation{Department of Physics, The University of Texas at Dallas, Richardson, Texas
75080-3021, USA}
\author{Xi-Wang Luo}
\affiliation{Department of Physics, The University of Texas at Dallas, Richardson, Texas
75080-3021, USA}
\author{Kuei Sun}
\affiliation{Department of Physics, The University of Texas at Dallas, Richardson, Texas
75080-3021, USA}
\author{Thomas Bersano}
\affiliation{Department of Physics and Astronomy, Washington State University, Pullman,
WA 99164, USA}
\author{Vandna Gokhroo}
\affiliation{Department of Physics and Astronomy, Washington State University, Pullman,
WA 99164, USA}
\author{Sean Mossman}
\affiliation{Department of Physics and Astronomy, Washington State University, Pullman,
WA 99164, USA}
\author{Peter Engels}
\affiliation{Department of Physics and Astronomy, Washington State University, Pullman,
WA 99164, USA}
\author{Chuanwei Zhang}
\thanks{Corresponding author. \\
Email: \href{mailto:chuanwei.zhang@utdallas.edu}{chuanwei.zhang@utdallas.edu}%
}
\affiliation{Department of Physics, The University of Texas at Dallas, Richardson, Texas
75080-3021, USA}
\title{Momentum-Space Josephson Effects}

\begin{abstract}
The Josephson effect is a prominent phenomenon of quantum supercurrents that
has been widely studied in superconductors and superfluids. Typical
Josephson junctions consist of two real-space superconductors (superfluids)
coupled through a weak tunneling barrier. Here we propose a momentum-space
Josephson junction in a spin-orbit coupled Bose-Einstein condensate, where
states with two different momenta are coupled through Raman-assisted
tunneling. We show that Josephson currents can be induced not only by
applying the equivalent of ``voltages'', but also by tuning tunneling
phases. Such \textit{tunneling-phase-driven Josephson junctions} in momentum
space are characterized through both full mean field analysis and a concise
two-level model, demonstrating the important role of interactions between
atoms. Our scheme provides a platform for experimentally realizing
momentum-space Josephson junctions and exploring their applications in
quantum-mechanical circuits.
\end{abstract}

\maketitle

\begin{figure}[t]
\centering
\includegraphics[width=0.48\textwidth]{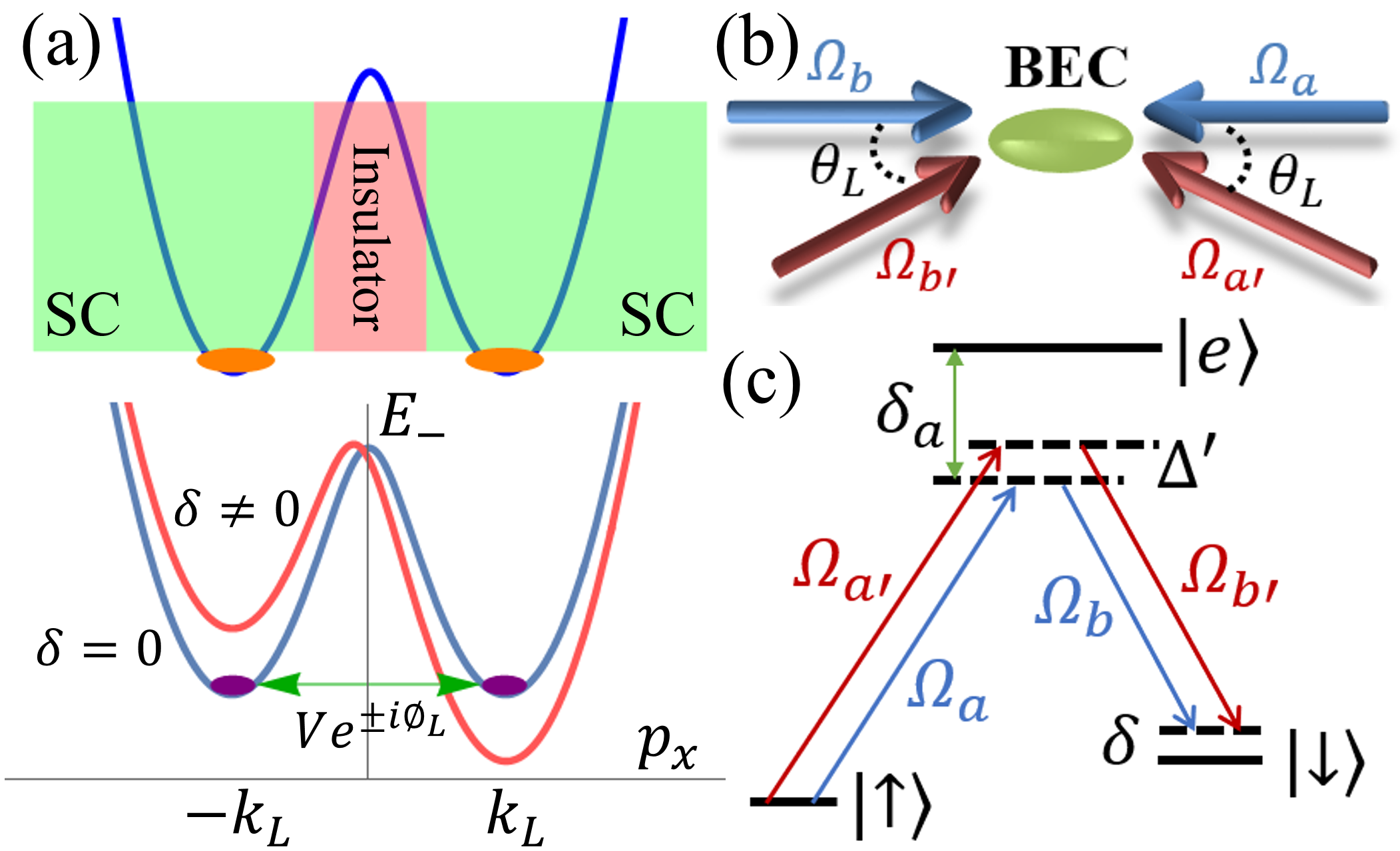}
\caption{(a) Illustration of conventional JJ for real-space superconductors
(top) versus MSJJ (bottom), where the double well band dispersion is
generated using a spin-orbit coupled BEC. (b,c) Experimental setup for
realizing a MSJJ. Two pairs of Raman lasers realize SOC (blue) and weak
coupling (red) between two band minima, respectively.}
\label{fig1}
\end{figure}

{\color{blue}\emph{Introduction}}. The Josephson effect \cite%
{Josephson1962,Josephson1974} is an intriguing quantum phenomenon of
supercurrents across a device known as a Josephson junction (JJ). A typical
JJ consists of two macroscopic quantum systems [e.g., superconductors,
superfluids, or Bose-Einstein condensates (BECs)] that are separated in real
or spin space and weakly coupled by quantum tunneling through a thin barrier
[Fig.~\ref{fig1}(a)] or by Rabi coupling between different spins. Because of
quantum tunneling of particles across the junction, JJs have found important
applications in quantum-mechanical circuits, such as SQUIDs \cite%
{Ryu2013,Makhlin2001}, superconducting qubits \cite%
{Martinis2001,Astafiev2006,Martinis2009,Paik2011}, and precision
measurements \cite{Makhlin2001}. In experiments, JJs have been widely
realized in solid state superconductors \cite{Giaever1960,Likharev1979},
superfluid Helium \cite%
{Pereverzev1997,Hoskinson2005,Wheatley1975,Leggett1975}, and recently, in
ultra-cold atomic gases \cite%
{Dalfovo1999,Smerzi1997,Raghavan1999,Williams1999,Ohberg1999,Cataliotti2001,Zibold2010,Albiez2005,Levy2007,Valtolina2015,Burchianti2017,Burchianti2}%
, where oscillating supercurrents were generated by applying a voltage drop
(or its equivalent) across JJs while maintaining a constant weak coupling (%
\textit{i.e.}, a.c.~Josephson effect \cite{Anderson1967}).

While JJs have been well studied in real space, a natural and important
question is whether Josephson effects can also be observed in momentum
space. In this paper, we address this question and propose a scheme for
realizing momentum-space JJs (MSJJs). In analogy to bosonic JJs in a
real-space double well \cite{Albiez2005,Levy2007}, a MSJJ may be realized
with a momentum-space double-well dispersion [see Fig. \ref{fig1}(a)], which
is an essential property of spin-orbit coupled systems \cite%
{Goldman2014,Lin2011}. Spin-orbit coupling (SOC) is ubiquitous in solid
state materials and has recently been realized experimentally in ultracold
atomic gases \cite%
{Lin2011,Zhang2012b,Qu2013a,Olson2014,Hamner2014,Wang2012,Cheuk2012,Williams2013,Lev,Jo,Huang2016, Meng2016,Pan2016}%
. In the presence of SOC, condensates at distinct band minima can be
considered as two distinct independent quantum systems. However, unlike
quantum tunneling between two wells in real space, two BECs at distinct
momenta are not directly coupled.

Here we propose a MSJJ facilitated by a tunable inter-well coupling in an
spin-orbit coupled BEC \cite{Galitski-review,Zhang2016}, where the coupling
is generated by an additional pair of counter-propagating Raman lasers. Such
Raman-assisted tunneling between two momentum states changes both the atomic
spin and momentum, and thus couples the condensates at the two band minima.
The SOC coupling strength dictates the height of the insulating barrier
while the Raman detuning serves as an effective voltage between the two band
minima. Suddenly changing the detuning (\textit{i.e.}, applying a voltage)
induces a coherent oscillation of the BECs between the two band minima (%
\textit{i.e.}, supercurrent oscillations), similar to traditional
a.c.~Josephson effects in superconductors. More interestingly, the phase of
the Raman-assisted tunneling between BECs at the two band minima is highly
tunable \cite{Spielman2015}, in contrast to real tunneling coefficients for
real-space JJs in superconductors \cite{Giaever1960,Likharev1979} and
double-well BECs \cite{Albiez2005,Levy2007}. We show that a sudden change of
the tunneling phase (while keeping the effective voltage unchanged) can also
induce Josephson effects of supercurrents, a phenomenon that we name as
\textquotedblleft \textit{tunneling-phase-driven JJ}\textquotedblright . We
focus on this new type of Josephson effect and study its properties through
both full mean-field simulation with the Gross-Pitaevskii equation (GPE)
\cite{Dalfovo1999,BEC-book} and the development of an effective two-level
model. Our results present rich physics in this system with different types
of supercurrent oscillations (Josephson, plasmonic \cite{Raghavan1999},
self-trapping \cite{Raghavan1999,Albiez2005}, etc.) and display the
important role of many-body interactions between atoms. Due to their
stability and high controllability, the proposed MSJJs and
tunneling-phase-driven JJs may have potential applications for building
novel quantum mechanical circuits.

{\color{blue}\emph{Experimental setup and theoretical modeling}}. We
consider a BEC confined in an elongated trap. Two internal states $%
\left\vert \uparrow \right\rangle $ and $\left\vert \downarrow \right\rangle
$ are coupled by two counter-propagating Raman lasers with Rabi frequencies $%
\Omega _{a}$ and $\Omega _{b}$, forming an effective one-dimensional (1D)
SOC dispersion relation along the $x$ direction [see Fig.~\ref{fig1}(b, c)].
Hereafter we choose recoil momentum $\hbar k_{R}$ and recoil energy $%
E_{R}=\hbar ^{2}k_{R}^{2}/2m$ for the Raman lasers as the units of momentum
and energy. Consequently, we have length and time in units of $2\pi /k_{R}$
and $\hbar /E_{R}$. The 1D SOC displays a double-well band dispersion in
momentum space with two band minima located at $\pm k_{L}=\pm \sqrt{1-\left(
\Omega /4\right) ^{2}}$, where $\Omega $ is the Raman coupling strength \cite%
{Li2012}. The tunneling between BECs at $\pm k_{L}$ requires simultaneous
change of spin and momentum, which can be realized using another independent
pair of Raman lasers $\Omega _{a^{\prime }}$ and $\Omega _{b^{\prime }}$
incident at an angle $\theta _{L}=\arccos \left( 1-k_{L}\right) $ to the $x$
axis [Fig.~\ref{fig1}(b)]. The frequencies of the pair $(a^{\prime
},b^{\prime })$ are shifted from those of the pair $(a,b)$ by $\Delta
^{\prime }\sim $ 100 MHz so that the interference between them is
negligible. The frequency difference between $a^{\prime }\ $and $b^{\prime }$
should match that between $a$ and $b$ to generate a time-independent
coupling.

Since only the $x$ direction is relevant for the SOC dynamics, the other two
directions can be integrated out, yielding an effective 1D system. The
dynamics of the system can be described by the GPE%
\begin{equation}
i\frac{\partial }{\partial t}\psi =(H_{0}+\frac{1}{2}\omega _{x}^{2}x^{2}+%
\frac{g}{2}|\psi |^{2})\psi
\end{equation}%
under the mean-field approximation, where $\psi =\left( \psi _{\uparrow
},\psi _{\downarrow }\right) ^{T}$ is the two component condensate
wavefunction normalized by the average particle number density $n=\int
dx\psi ^{\dagger }\psi $, $\omega _{x}$ represents trapping frequency of
harmonic trap. For a typical $^{87}$Rb BEC, the effective density
interaction $ng\sim 0.1$ with $\sim 10^{4}$ atoms (see "Experimental
consideration" section) and the spin interaction is negligible. The Raman
coupling does not affect atomic interactions. The single particle
Hamiltonian can be written as~\cite{Supp,Brion2007}
\begin{equation}
H_{0}=\left(
\begin{array}{cc}
(p_{x}-1)^{2}-\frac{\delta }{2} & \frac{\Omega }{2}+e^{i\phi _{L}}\Omega
_{L}e^{2ik_{L}x} \\
\frac{\Omega }{2}+e^{-i\phi _{L}}\Omega _{L}e^{-2ik_{L}x} & (p_{x}+1)^{2}+%
\frac{\delta }{2}%
\end{array}%
\right) ,  \label{SH}
\end{equation}%
where $\Omega _{L}$ is the coupling strength generated by the tunneling
lasers, $\phi _{L}$ is the relative phase between the two Raman couplings,
and $\delta $ is the detuning.

\begin{figure}[t]
\centering
\includegraphics[width=0.48\textwidth]{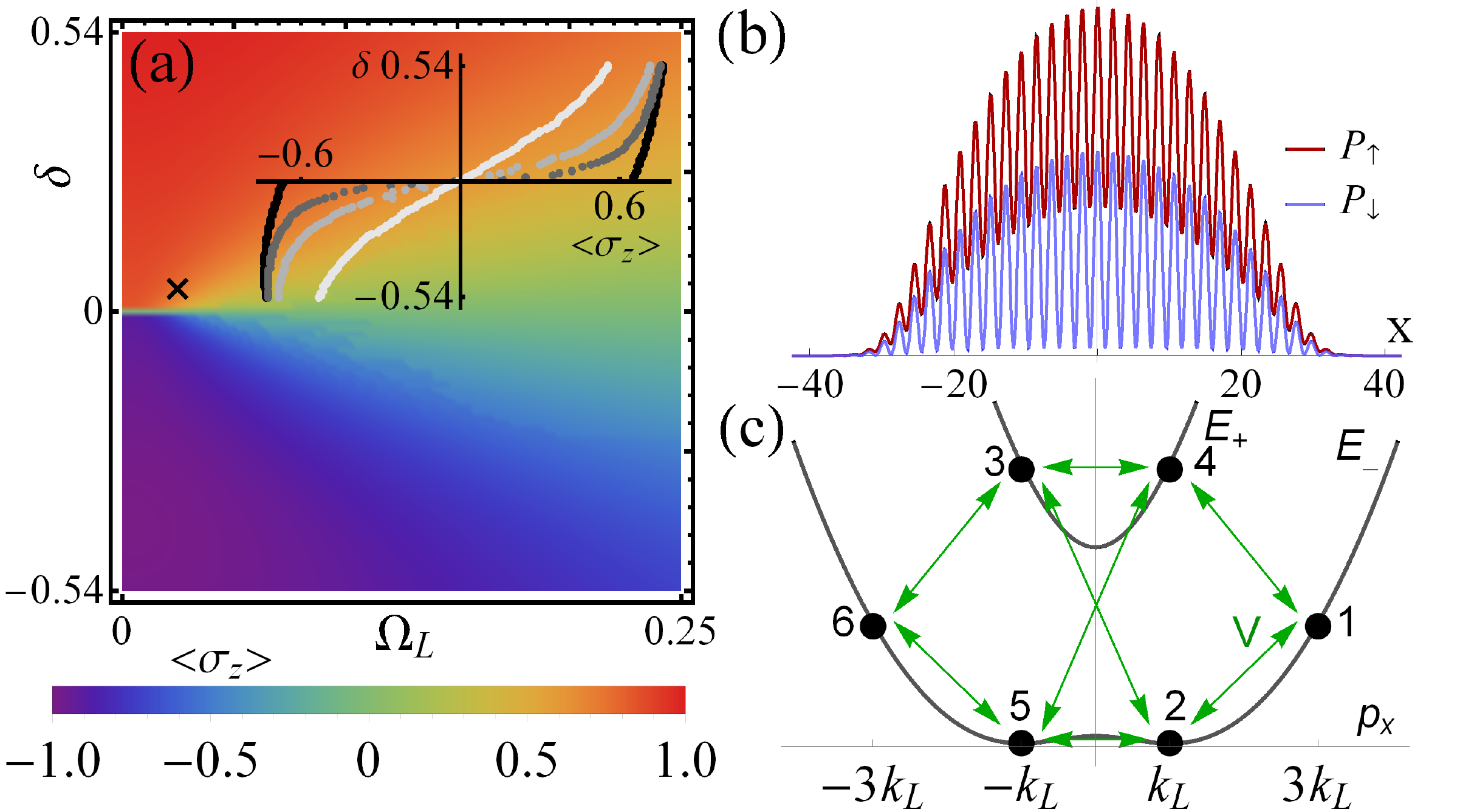}
\caption{(a) Ground state phase diagram, where $\Omega =2.7$, $\protect\phi %
_{L}=0$ and $ng=0.07$. The inset shows the first-order phase transition for
small $\Omega _{L}$. Black, dark gray, light gray and white lines correspond
to $\Omega _{L}=0.01$, $0.1$, $0.2$ and $0.5$ respectively. (b) Real space
density modulation for the ground state with parameters $\protect\delta %
=0.054$ and $\Omega _{L}=0.015$ as denoted by the black cross in (a). (c)
Illustration of induced couplings between six most relevant momentum states.}
\label{fig2}
\end{figure}

The ground state of the BEC is obtained from the imaginary time evolution of
the GPE \cite{Supp,Bader2013} using a time-split-operator method, resulting
in the phase diagram shown in Fig.~\ref{fig2}(a) in the $\Omega _{L}$-$%
\delta $ plane, where the color represents spin polarization $\langle \sigma
_{z}\rangle $. For weak $\Omega _{L}$, interactions lock the condensate to
one momentum minimum, yielding a plane-wave phase at large detunings. There
is a first-order phase transition [black line in the inset of Fig~\ref{fig2}%
(a)] when $\delta $ crosses $0$. With increasing $\Omega _{L}$, the
single-particle coupling dominates over the interaction, hence the ground
state is in a stripe-like phase with a real-space density modulation [Fig.~%
\ref{fig2}(b)], and $\langle \sigma _{z}\rangle $ varies continuously and
smoothly with respect to $\delta $ (white line in the inset of Fig. \ref%
{fig2}(a)). While a supersolid stripe phase is defined through spontaneous
breaking of both continuous translational and gauge symmetries \cite%
{Martone2014,Martone2016}, here continuous translational symmetry is
synthetically broken by the periodic potential $e^{2ik_{L}x}$. Nevertheless,
the ground state is the superposition of two band minima, similar to an
authentic stripe phase induced by interactions.

The additional Raman lasers $\Omega _{L}$ couple not only the two band
minima, but also other states from both lower and upper bands. The six most
relevant momentum states $\psi _{i}$ are shown in Fig.~\ref{fig2}(c).
Expanding the wavefunction $\psi =\sum_{i=1}^{6}C_{i}\psi _{i}$ in this
six-state basis, we obtain a $6\times 6$ effective Hamiltonian \cite{Supp}.
The direct coupling between the two band minima at $2$ and $5$ is $%
-V_{0}e^{\mp i\phi _{L}}$ with $V_{0}=\frac{1}{2}\Omega _{L}(1+k_{L})$,
while the couplings with other neighboring high-energy states are $-\sqrt{%
\frac{1-k_{L}}{2}}e^{\pm i\phi _{L}}$ and $\frac{1}{2}\sqrt{1-k_{L}^{2}}%
e^{\mp i\phi _{L}} $, which approach $0$ when $k_{L}\rightarrow 1$, leaving $%
V_{0}$ as the dominant tunneling term. We focus on the region $\Omega
_{L}\ll \Omega $ to avoid significant modification of the original SOC band
dispersion and also for the observation of Josephson effects with weak
tunneling.

\begin{figure}[t]
\centering
\includegraphics[width=0.48\textwidth]{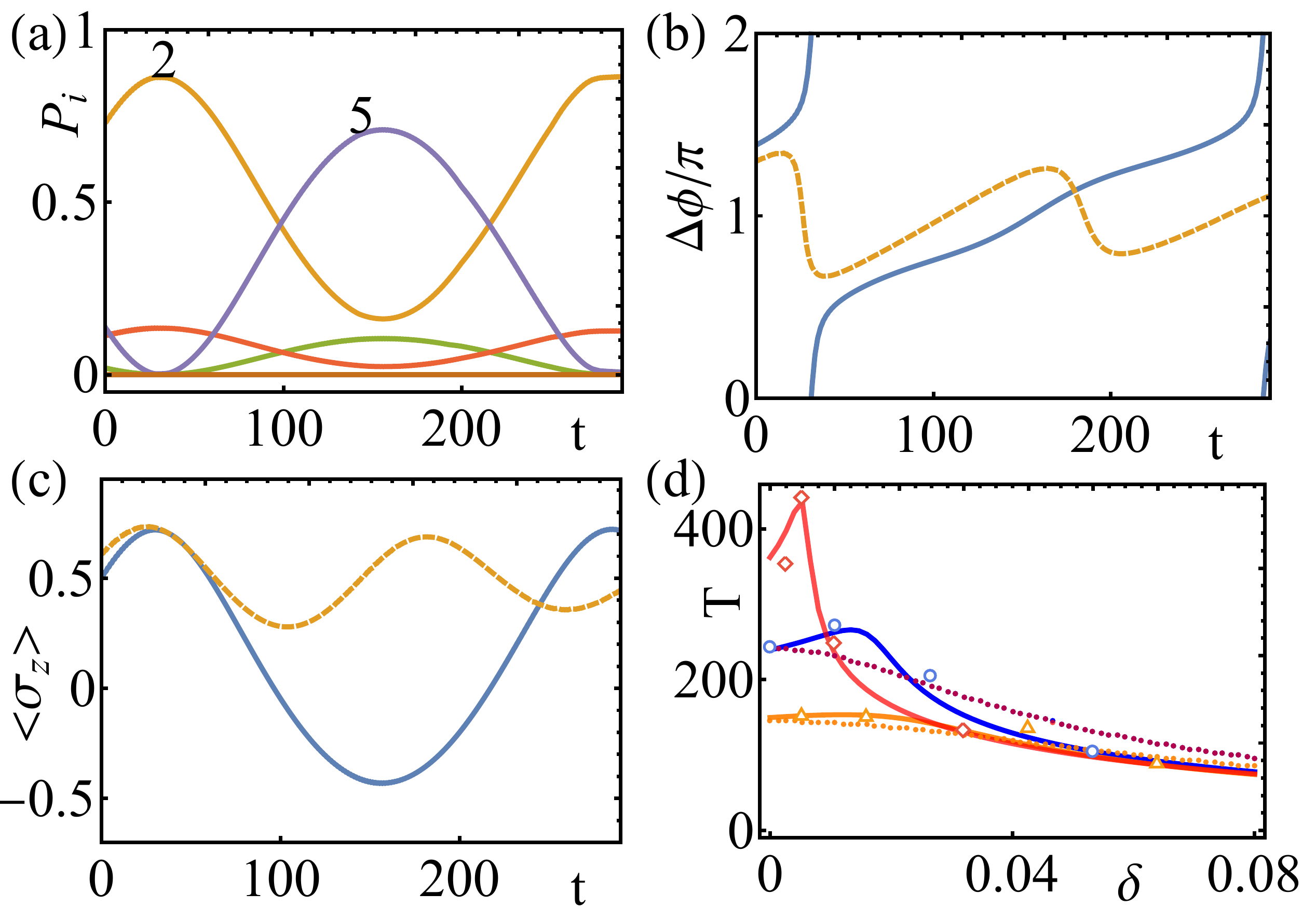}
\caption{(a) Time evolution of the populations at different momentum states
for tunneling-phase-driven MSJJ for $\Omega =2.7$, $\protect\delta =0.014$, $%
\Omega _{L}=0.015$ and $ng=0.07$. (b) and (c): Evolutions of phase
difference (b) and polarization (c) for Josephson oscillation (solid blue)
and plasma oscillation (dashed orange). (d) Oscillation period $T$ versus $%
\protect\delta $ when $\protect\phi _{L}$ is quenched from $\protect\phi %
_{L0}$ to $0$ at $\Omega =2.7$. Circles are results from the GPE simulation,
while solid ($ng=0.07$) and dashed (single particle) lines are from the
two-level model. Different colors correspond to different parameter sets: $%
\Omega _{L}=0.015$, $\protect\phi _{L0}=0.4\protect\pi $ (blue); $\Omega
_{L}=0.025$, $\protect\phi _{L0}=0.4\protect\pi $ (orange); and $\Omega
_{L}=0.015$, $\protect\phi _{L0}=0.2\protect\pi $ (red). Blue and red dashed
lines overlap (purple) since $T$ is independent of $\protect\phi _{L0}$ for
the single particle case.}
\label{fig3}
\end{figure}

{\color{blue}\emph{Tunneling-phase-driven MSJJ}}. In an a.c. JJ, a suddenly
applied voltage can induce an oscillation of supercurrents between two
superconductors. In our system, BECs at the two band minima marked $2$ and $%
5 $ are considered as two superfluids and the detuning between them
corresponds to a voltage. A sudden change of $\delta $ induces an
oscillation of the BEC between the two minima, yielding a MSJJ whose
properties are described in the supplementary materials \cite{Supp}. Here we
focus on the relative phase $\phi _{L}$ for the tunneling element between $2$
and $5$, which is highly tunable in experiments \cite{Spielman2015}. In
contrast, such tunneling is a real number for a real space JJ between two
superconductors or double well BECs. A sudden change of the phase $\phi _{L}$
(keeping $\delta $ constant) can induce a different type of Josephson
effect, \textit{i.e.}, \textit{tunneling-phase-driven JJ}.

In Figs.~\ref{fig3}(a)--(c) we show dynamics from simulations of the GPE
with a sudden change of the phase $\phi _{L}$ from an initial $\phi _{L0}$
to $\phi _{Lf}=0$. In panel (a) we plot the population $P_{i}\left( t\right)
$ at each momentum state for $\phi _{L0}=0.4\pi $. Clearly only the states $%
2 $ and $5$ at the two band minima are largely populated while all other
states can be neglected due to their small initial populations, weak
coupling to states $2$ and $5$, and high energies. Panel (b) shows the
relative phase between BECs in states $2$ and $5$. For $\phi _{L0}=0.4\pi $
(blue solid line), the phase varies through $[0,2\pi )$, representing a
Josephson type of oscillation; while for $\phi _{L0}=0.3\pi $ (yellow dashed
line), the phase oscillates in a small range, showing a plasma oscillation.
The polarization $\langle \sigma _{z}\rangle $ exhibits sinusoidal
oscillations for both cases [panel (c)].

Because the population of the BEC stays mainly at the two band minima $2$
and $5$, we can neglect the other states to derive an effective two-level
model, yielding an equation of motion \cite{Supp, Li2012}
\begin{equation}
i\partial _{t}\left(
\begin{array}{c}
C_{2} \\
C_{5}%
\end{array}%
\right) =\left( H_{0}^{\mathrm{eff}}+H_{\mathrm{I}}^{\mathrm{eff}}\right)
\left(
\begin{array}{c}
C_{2} \\
C_{5}%
\end{array}%
\right) ,  \label{2LevelRabi}
\end{equation}%
where $H_{0}^{\mathrm{eff}}=\left(
\begin{array}{cc}
-k_{L}\delta & -V_{0}e^{-i\phi _{L}} \\
-V_{0}e^{i\phi _{L}} & k_{L}\delta%
\end{array}%
\right) $ is the effective single-particle Hamiltonian, and $H_{\mathrm{I}}^{%
\mathrm{eff}}=2g_{G}\left(
\begin{array}{cc}
\left\vert C_{5}\right\vert ^{2} & 0 \\
0 & \left\vert C_{2}\right\vert ^{2}%
\end{array}%
\right) $ is the effective interaction term obtained through a variational
approximation of the GPE. Generally, $g_{G}$ depends on $\left\vert
C_{2}\right\vert ^{2}\left\vert C_{5}\right\vert ^{2}$ but is approximately
a constant when the interaction strength is weak compared to $E_{R}$,
yielding $g_{G}=ng(1-k_{L}^{2})$. Note that the coupling phase $\phi _{L}$
in Eq. (\ref{2LevelRabi}) can be incorporated into the relative phase
between $C_{2}$ and $C_{5}$ through a simple phase transformation, therefore
the quench of $\phi _{L}$ is mathematically equivalent to a quench of the
relative phase between condensates at two minima $\left( 2,5\right) $,
although the latter is experimentally impractical.

When the coupling $V_{0}$ is strong, the dynamics of the BEC are governed by
single particle physics, yielding a linear Rabi oscillation with period $%
T=\pi /\omega $, where the Rabi frequency $\omega =\sqrt{(k_{L}\delta
)^{2}+|V_{0}|^{2}}$. Such a simple formula for the period does not apply
when the tunneling $V_{0}$ is comparable to or weaker than the
inter-particle interactions, although the two-level model still agrees
reasonably well with the GPE simulations, as shown in Fig.~\ref{fig3}(d). We
see that the period is similar for interacting and single-particle cases for
a large coupling $\Omega _{L}=0.025$, but shows strong deviations [see the
sharp peak for the solid red line in Fig.~\ref{fig3}(d)] from the single
particle curve for $\Omega _{L}=0.015$. For a very large detuning $\delta $
(i.e., voltage), all $T$ collapse to the same line as the single particle
case, as expected.

\begin{figure}[t]
\centering
\includegraphics[width=0.48\textwidth]{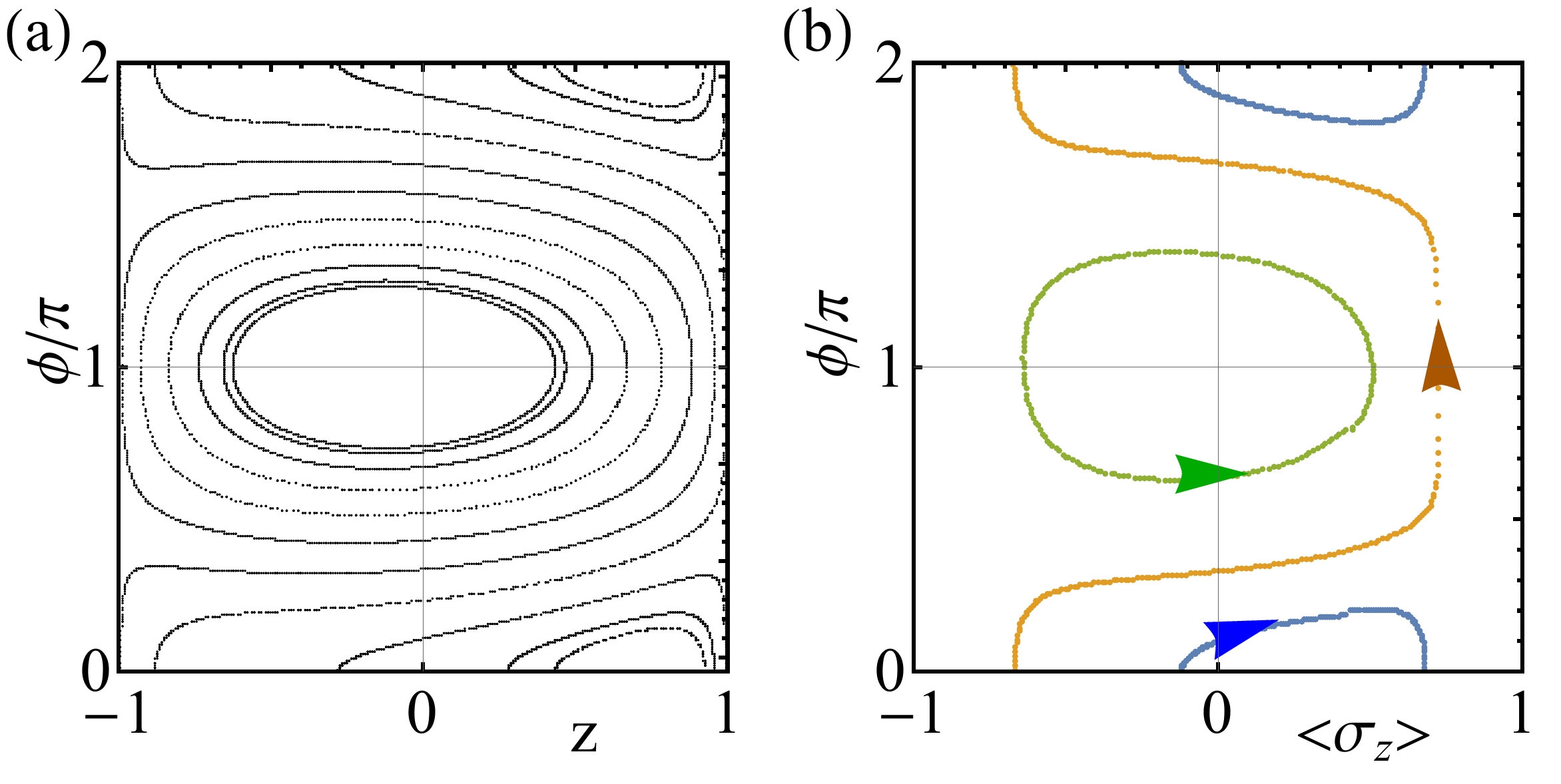}
\caption{(a) Classical trajectories in $z$-$\protect\phi $ plane for $0<%
\protect\phi _{L0}\leq \protect\pi $, with initial value of $z$ at $0.434$.
(b) Same as (a) but generated through the GPE simulation. Parameters are $%
ng=0.07$, $\protect\delta =0.008$ (corresponding to initial polarization $%
0.434$), $\Omega =2.7$ and $\Omega _{L}=0.015$. The three colors correspond
to $\protect\phi _{L0}=0.2$ (blue), $0.4$ (orange), $0.8$ (green),
respectively. The arrows denote the direction of each trajectory.}
\label{fig4}
\end{figure}

In the two-level approximation, we can choose the normalization $\left\vert
C_{2}\right\vert ^{2}+\left\vert C_{5}\right\vert ^{2}=1$, and recast the
equation of motion (\ref{2LevelRabi}) as \cite{Supp}
\begin{eqnarray}
\partial _{t}z &=&-\sqrt{1-z^{2}}\sin \left( \phi -\phi _{Lf}\right) ,
\label{EOM1} \\
\partial _{t}\phi &=&\frac{g_{G}}{V_{0}}z+\frac{z}{\sqrt{1-z^{2}}}\cos
\left( \phi -\phi _{Lf}\right) +\frac{k_{L}\delta }{V_{0}},  \label{EOM2}
\end{eqnarray}%
using the population difference $z=(N_{2}-N_{5})/N$ and relative phase $\phi
=\theta _{2}-\theta _{5}$, where $N_{i}$ and $\theta _{i}$ are defined
through $C_{2}=\sqrt{N_{2}}e^{i\theta _{2}}$ and $C_{5}=\sqrt{N_{5}}%
e^{i\theta _{5}}$. These two classical equations characterize the essential
dynamics of MSJJs.

Fig.~\ref{fig4}(a) shows how the initial value $\phi _{L0}$ affects the
dynamics. For a relatively small $\phi _{L0}$, the classical trajectory is a
closed loop around a fixed point with a small amplitude of $z$ and a
confined range of phase change $\Delta \phi $, showing a plasma oscillation
\cite{Raghavan1999}. With increasing $\phi _{L0}$, the amplitudes for both $%
\phi $ and $z$ increase. Beyond a critical $\phi _{L0}$, $\phi $ varies
through $[0,2\pi )$, showing a Josephson oscillation. The system returns to
plasma oscillation around another fixed point when $\phi _{L0}$ exceeds
another critical point. These classical trajectories from the two-level
model agree with those from the GPE simulations in Fig.~\ref{fig4}(b). Note
that the trajectories around two fixed points have opposite directions. In
the single-particle case, these two fixed points correspond to two opposite
Zeeman fields for spin precession of the Rabi oscillation \cite{Supp}.

\begin{figure}[t]
\centering
\includegraphics[width=0.48\textwidth]{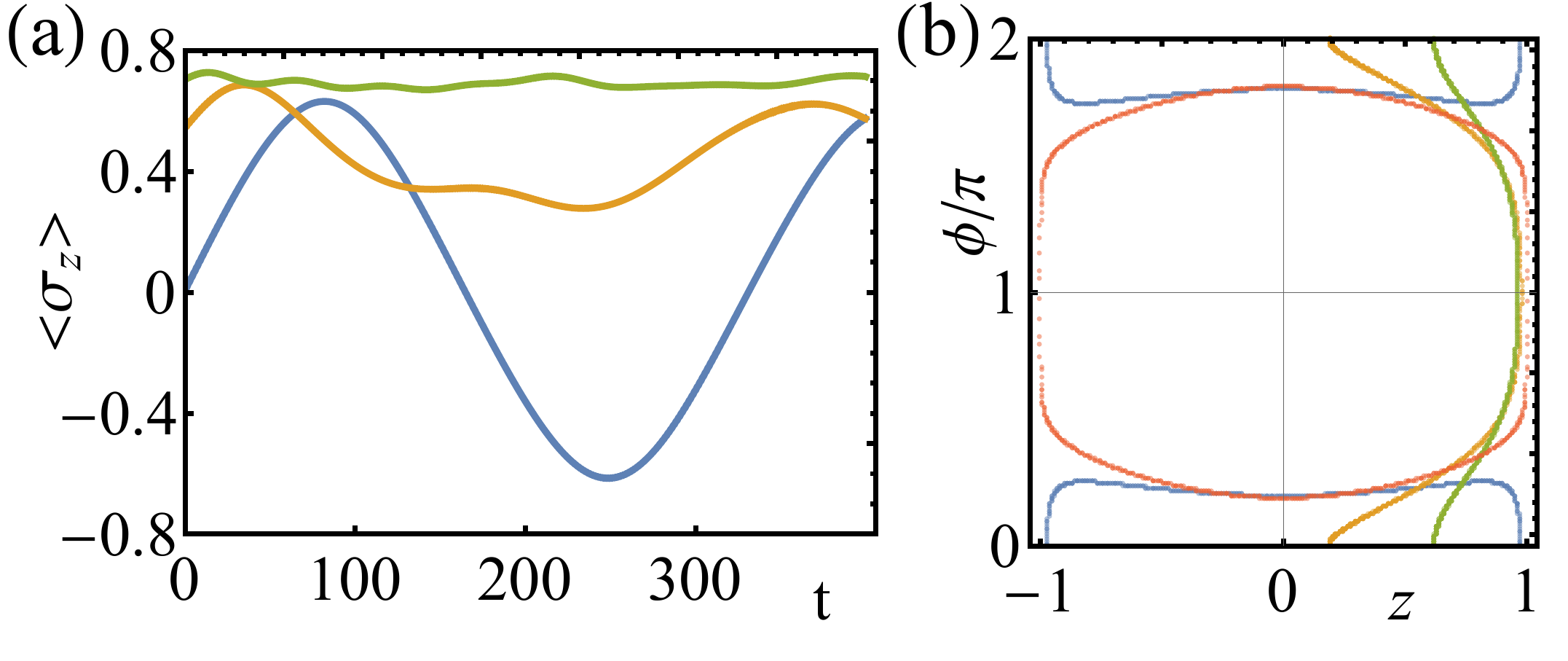}
\caption{(a) Self-trapping effects from the GPE simulation. The curves
correspond to $ng=0.07$ (blue), $0.14$ (orange) and $0.35$ (green), for $%
\Omega =2.7$, $\Omega _{L}=0.03$, $\protect\phi _{L0}=0.2\protect\pi $ and $%
\protect\delta =0$. (b) Classical phase-plane from the two-level model. The
colors are the same as in (a) except $ng=0.1$ for the red curve.}
\label{fig5}
\end{figure}

Strong interaction between atoms can dramatically change the BEC dynamics
and lead to a self-trapping effect \cite{Raghavan1999,Albiez2005}, where the
oscillation amplitude of $z$ is strongly suppressed. We consider a symmetric
oscillation with $\delta =0$. For a weak interaction of $ng=0.07$, the
oscillation of $\left\langle \sigma _{z}\right\rangle $ shows a perfect
sinusoidal pattern (blue line), as seen by the blue line Fig.~\ref{fig5}(a)
obtained from the GPE simulation. When the interaction is doubled $ng=0.14$,
the oscillation amplitude is reduced and the average $\left\langle \sigma
_{z}\right\rangle $ in one period changes from $0$ to a finite value (orange
line). For a larger but still practicable interaction of $ng=0.35$, the
oscillatory behavior disappears and the condensate is locked at the initial
band minimum because of strong density interaction. Such nonlinear
self-trapping effects can also be captured in the classical trajectories in
the two-level model [Fig.~\ref{fig5} (b)]. With increasing $ng$, the initial
plasma oscillation with a large variation of $z$ becomes the self-trapped
Josephson oscillation with a small $z$ change. 

{\color{blue}\emph{Experimental consideration}}. The periodic density
modulation for the stripe-like ground state can be measured using Bragg
scattering, similar to the recent experiments for observing supersolid
stripe phases \cite{JunRuLi2017}. Consider a ${}^{87}$Rb BEC confined in a
quasi-$1$D harmonic trap. The Raman lasers for generating SOC are incident
at 45$^{\circ }$ with the $x$ axis, yielding an effective wavevector $k_{R}=%
\frac{2\pi }{\sqrt{2}\lambda }\ $with $\lambda =784$ nm. The corresponding
recoil energy $E_{R}=2\pi \hbar \times 1.8$ kHz, thus the time and length
units are $\hbar /E_{R}=0.088$ ms and $2\pi /k_{R}=1109$ $\mu m$,
respectively. The Raman coupling strength for SOC $\Omega =2.7E_{R}$, thus $%
k_{L}=0.738k_{R}$ and the second pair of Raman lasers should be incident at
an angle $\theta _{L}=58.6^{\circ }$ with respect to the $x$ axis. The
s-wave scattering length of ${}^{87}$Rb is $a_{s}=100.86a_{0}$, where $a_{0}$
is the Bohr radius. Considering a particle number $10^{4}$ to $10^{6}$ and
typical trapping frequencies $\omega _{x}\sim 2\pi \times 5$ Hz and $\omega
_{y}=\omega _{z}\sim 2\pi \times 75$ Hz, one has the average particle
density $n\sim 10^{13}$ to $10^{14}~\text{cm}^{-3}$ under Thomas-Fermi
approximation \cite{BEC-book}. The effective interaction strength can be
evaluated through $ng=4\pi \hbar ^{2}a_{s}n/m\sim 0.07$ to $0.48~E_{R}$,
resulting in the time period $T\sim $ 10 ms for tunneling-phase-driven
Josephson oscillations [Fig.~\ref{fig3}(d)].

{\color{blue}\emph{Discussion and Conclusion}}. Our two major proposed
concepts, momentum-space JJ and tunneling-phase-driven JJ, may also be
realized in other physical systems where a double well band dispersion with
two almost degenerate local band minima can be generated to ensure the long
life time of the BEC at different momenta \cite{An2017}. For instance, the
double-well band dispersion may be realized in optical superlattices with
Raman assisted tunneling \cite{Li2016}, where two momentum minima can be
coupled with additional Raman transitions. The double-well band dispersion
can be generalized to triple-well or even more multiple-degenerate momentum
states, and the coupling between neighboring minima may form a
momentum-space optical lattice \cite{An2017}, which can be considered as a
Josephson junction array \cite{Cataliotti2001} in momentum space. The linear
momentum discussed here can be generalized to orbital angular momentum
(OAM), and an OAM-space JJ may be realized for a BEC on a ring utilizing
recent proposals for spin-OAM coupling \cite{Sun2015, DeMarco2015, Qu2015}
for cold atoms. The discreteness of OAM states may induce interesting
Josephson effects that are different from those in continuous real or
momentum space. Finally, although absent in solid-state superconductors, the
proposed tunneling-phase-induced JJ may be realized in real-space optical
superlattices with Raman assisted tunneling \cite{Li2016}, where the phase
for the Raman tunneling may also be tuned.

In conclusion, we propose a new category of Josephson effects in momentum
space, which can be built in a spin-orbit coupled BEC. In addition to
traditional voltage-driven Josephson effects, we introduce quenching of the
tunneling phase as a novel driving mechanism. Our work may motivate further
experimental and theoretical works for studying MSJJs and provides a
platform for exploring their applications in building novel quantum
mechanical circuits.

\begin{acknowledgments}
\textbf{Acknowledgements}: This work was supported by Air Force Office of
Scientific Research (FA9550-16-1-0387), National Science Foundation
(PHY-1505496), and Army Research Office (W911NF-17-1-0128). P. E.
acknowledges funding from National Science Foundation (PHY-1607495).
\end{acknowledgments}

\newpage \clearpage
\onecolumngrid
\appendix

\section{Supplementary materials}
\subsection{Single particle Hamiltonian}

Here we derive the single-particle Hamiltonian of Eq.~(2) in the main text
using adiabatic elimination and the rotating wave approximation. The atomic
lambda system consists of two atomic hyperfine ground states, denoted as $%
\left\vert \uparrow \right\rangle $, $\left\vert \downarrow \right\rangle $,
and an excited state $|e\rangle $. The system is driven by two pairs of
off-resonance lasers $(\omega _{a},\omega _{b})$ and $(\omega _{a^{\prime
}},\omega _{b^{\prime }})$, as illustrated in Fig.~1(b, c) (main text). The
frequency $\omega _{a^{\prime }}$ ($\omega _{b^{\prime }}$) is shifted by $%
\Delta ^{\prime }\sim 100$ MHz from $\omega _{a}$ ($\omega _{b}$) so that
the interference effects can be neglected and the two pairs can be treated
individually. The frequency of each Raman laser satisfies $\delta
_{a}=\omega _{e\uparrow }-\omega _{a}$ and $\delta _{b}=\omega _{e\downarrow
}-\omega _{b}$, where $\omega _{e\uparrow }$($\omega _{e\downarrow }$) is
the energy difference between $|e\rangle $ and $\left\vert \uparrow
\right\rangle $($\left\vert \downarrow \right\rangle $). In the rotating
frame defined by the transition $\left(
\begin{array}{c}
|\uparrow \rangle \\
|\downarrow \rangle \\
|e\rangle%
\end{array}%
\right) \rightarrow e^{i\widehat{R_{1}}}\left(
\begin{array}{c}
|\uparrow \rangle \\
|\downarrow \rangle \\
|e\rangle%
\end{array}%
\right) $, where $\widehat{R_{1}}=\text{diag}(\delta /2,\omega _{\uparrow
}-\omega _{\downarrow }-\delta /2,\omega _{\uparrow }-\omega _{e}+\Delta )$
with $\delta =\delta _{a}-\delta _{b}$ and $\Delta =\left( \delta
_{a}+\delta _{b}\right) /2$, the laser-atom interaction Hamiltonian becomes
\cite{Brion2007}
\begin{equation*}
H_{\text{LA}}=\left(
\begin{array}{ccc}
-\frac{\delta }{2} & 0 & \frac{\Omega _{a}^{\ast }}{2}\left( 1+\frac{\Omega
_{a^{\prime }}^{\ast }}{\Omega _{a}^{\ast }}e^{i(\omega _{a^{\prime
}}-\omega _{a})t}\right) \\
0 & \frac{\delta }{2} & \frac{\Omega _{b}^{\ast }}{2}\left( 1+\frac{\Omega
_{b^{\prime }}^{\ast }}{\Omega _{b}^{\ast }}e^{i(\omega _{b^{\prime
}}-\omega _{b})t}\right) \\
\frac{\Omega _{a}}{2}\left( 1+\frac{\Omega _{a^{\prime }}}{\Omega _{a}}%
e^{i(\omega _{a}-\omega _{a^{\prime }})t}\right) & \frac{\Omega _{b}}{2}%
\left( 1+\frac{\Omega _{b^{\prime }}}{\Omega _{b}}e^{i(\omega _{b}-\omega
_{b^{\prime }})t}\right) & \Delta%
\end{array}%
\right) .
\end{equation*}%
Here $\Omega _{i}$ is the Rabi coupling strength. Since $\Delta \gg |\delta
|,$ $\left\vert \Omega _{i}\right\vert $, the excited state can be adiabatic
eliminated, yielding
\begin{equation*}
H_{\text{LA}}^{\mathrm{eff}}=\left(
\begin{array}{cc}
-\delta /2 & \frac{\tilde{\Omega}_{a}\tilde{\Omega}_{b}^{\ast }}{2\Delta }
\\
\frac{\tilde{\Omega}_{a}^{\ast }\tilde{\Omega}_{b}}{2\Delta } & \delta /2%
\end{array}%
\right) ,
\end{equation*}%
where $\tilde{\Omega}_{a}=\Omega _{a}\left( 1+\frac{\Omega _{a^{\prime }}}{%
\Omega _{a}}e^{i\tilde{\omega}_{a}t}\right) $ and $\tilde{\Omega}_{b}=\Omega
_{b}\left( 1+\frac{\Omega _{b^{\prime }}}{\Omega _{b}}e^{i\tilde{\omega}%
_{b}t}\right) $ with $\tilde{\omega}_{i}=\omega _{i^{\prime }}-\omega _{i}$.
Taking $\tilde{\omega}_{a}=\tilde{\omega}_{b}$ (i.e., the two pairs of Raman
lasers have the same frequency difference) and neglecting fast time
modulating terms, the Raman coupling becomes $\tilde{\Omega}_{a}\tilde{\Omega%
}_{b}^{\ast }=\Omega _{a}\Omega _{b}^{\ast }+\Omega _{a^{\prime }}\Omega
_{b^{\prime }}^{\ast }$, which is a direct summation of Raman couplings for
each laser pair. Considering the laser configuration in Fig.~1 of the main
text, we have
\begin{equation*}
H_{V}=\frac{(\hbar k_x)^{2}}{2m}+\frac{1}{2}\left(
\begin{array}{cc}
-\delta & \Omega e^{-2ik_{R}x}+2e^{i\phi _{L}}\Omega
_{L}e^{-2i(k_{R}-k_{L})x} \\
\Omega e^{2ik_{R}x}+2e^{-i\phi _{L}}\Omega _{L}e^{2i(k_{R}-k_{L})x} & \delta%
\end{array}%
\right),
\end{equation*}%
where $\phi _{L}$ is the phase difference between the two Raman couplings
and $k_{L}<k_{R}$ since the primed lasers are injected at an angle $\theta
_{L}$. A standard unitary transformation of the spatially dependent phases
yields the single particle Hamiltonian
\begin{equation*}
H_{0}=\frac{1}{2}\left(
\begin{array}{cc}
\frac{\hbar^2}{m}(k_{x}-k_{R})^{2}-\delta & \Omega +e^{i\phi _{L}}2\Omega
_{L}e^{2ik_{L}x} \\
\Omega +e^{-i\phi _{L}}2\Omega _{L}e^{-2ik_{L}x} & \frac{\hbar^2}{m}%
(k_{x}+k_{R})^{2}+\delta%
\end{array}%
\right).
\end{equation*}

\subsection{2-level approximation and Rabi oscillation}

The second pair of Raman lasers can induce coupling between the $k$ and $%
k\pm 2k_{L} $ states with different spins in the SOC picture. Taking only
the $6$ most relevant neighboring states around the two band minima, the
single particle Hamiltonian can be projected to
\begin{equation*}
H=\left(
\begin{array}{cccccc}
(-1+3k_{L})^{2} & 0 & 0 & \Omega _{L}e^{i\phi _{L}} & 0 & 0 \\
0 & (-1+k_{L})^{2} & 0 & \Omega /2 & \Omega _{L}e^{i\phi _{L}} & 0 \\
0 & 0 & (-1-k_{L})^{2} & 0 & \Omega /2 & \Omega _{L}e^{i\phi _{L}} \\
\Omega _{L}e^{-i\phi _{L}} & \Omega /2 & 0 & (1+k_{L})^{2} & 0 & 0 \\
0 & \Omega _{L}e^{-i\phi _{L}} & \Omega /2 & 0 & (1-k_{L})^{2} & 0 \\
0 & 0 & \Omega _{L}e^{-i\phi _{L}} & 0 & 0 & (1-3k_{L})^{2}%
\end{array}%
\right) ,
\end{equation*}%
where we have set $k_{R}=1$ and rearranged the order to put states with the
same spin together. Using the relation ${k_{L}}=\sqrt{1-{{\left( {\frac{%
\Omega }{4}}\right) }^{2}}}$, we can diagonalize the Hamiltonian with $%
\Omega _{L}=0$, yielding the bare Hamiltonian $H_{B}=\text{diag}\left(
(1-3k_{L})^{2},-1+k_{L}^{2},3+k_{L}^{2},3+k_{L}^{2},-1+k_{L}^{2},(1-3k_{L})^{2}\right)
$ for the SOC band. The corresponding states are labeled from $\left\vert
1\right\rangle $ to $\left\vert 6\right\rangle $ in Fig. 2(c). In this new
basis, the total Hamiltonian can be rewritten as $H_{B}+\Omega _{L}H^{\prime
}$ with
\begin{eqnarray*}
H^{\prime }= \left(
\begin{array}{cccccc}
0 & \sqrt{\frac{1-k_{L}}{2}}e^{i\phi _{L}} & 0 & \sqrt{\frac{1+k_{L}}{2}}%
e^{i\phi _{L}} & 0 & 0 \\
\sqrt{\frac{1-k_{L}}{2}}e^{-i\phi _{L}} & 0 & -\frac{1}{2}\sqrt{1-k_{L}^{2}}%
e^{i\phi _{L}} & 0 & -\frac{1}{2}(1+k_{L})e^{i\phi _{L}} & 0 \\
0 & -\frac{1}{2}\sqrt{1-k_{L}^{2}}e^{-i\phi _{L}} & 0 & -\frac{1}{2}%
(-1+k_{L})e^{-i\phi _{L}} & 0 & \sqrt{\frac{1+k_{L}}{2}}e^{i\phi _{L}} \\
\sqrt{\frac{1+k_{L}}{2}}e^{-i\phi _{L}} & 0 & -\frac{1}{2}(-1+k_{L})e^{i\phi
_{L}} & 0 & \frac{1}{2}\sqrt{1-k_{L}^{2}}e^{i\phi _{L}} & 0 \\
0 & -\frac{1}{2}(1+k_{L})e^{-i\phi _{L}} & 0 & \frac{1}{2}\sqrt{1-k_{L}^{2}}%
e^{-i\phi _{L}} & 0 & -\sqrt{\frac{1-k_{L}}{2}}e^{i\phi _{L}} \\
0 & 0 & \sqrt{\frac{1+k_{L}}{2}}e^{-i\phi _{L}} & 0 & -\sqrt{\frac{1-k_{L}}{2%
}}e^{-i\phi _{L}} & 0%
\end{array}%
\right). \\
&&
\end{eqnarray*}%
Clearly, the coupling between the two band minima 2 and 5 with energy $%
-1+k_{L}^{2}$ is $V=-\frac{\Omega _{L}}{2}(1+k_{L})e^{\mp i\phi _{L}}$. Note
that, we only compute the coupling strength to the first-order of $\Omega_L$%
, and one may come to the same conclusion even using a minimal four-level
model (two minima and corresponding upper band states).

We neglect other state populations and thus project the Hamiltonian onto a
two-level model in Eq.(3). Without interaction, the single particle
Hamiltonian in this $2$-level approximation is rewritten as (without loss of
generality, we set $\phi _{L}=0$)
\begin{equation*}
H=-k_{L}\delta {\tau _{z}}-V_{0}{\tau _{x}}=\omega (-\cos \alpha ~{\tau _{z}}%
+\sin \alpha ~{\tau _{x}}),
\end{equation*}%
where $\tan \alpha =-V_{0}/(k_{L}\delta )$ and $\{{\tau }\}$ are Pauli
matrices. The oscillation of any state driven by this Hamiltonian can be
calculated using the time-evolution operator,
\begin{equation*}
{e^{-iHt}}=\left( {%
\begin{array}{cc}
{\cos \omega t+i\cos \alpha \sin \omega t} & {\ -i\sin \alpha \sin \omega t}
\\
{\ -i\sin \alpha \sin \omega t} & {\cos \omega t-i\cos \alpha \sin \omega t}%
\end{array}%
}\right) .
\end{equation*}

For a general state at $t=0$ (an initial state can be analytically obtained
using variational analysis and is discussed in the following section),
\begin{equation*}
\psi (0)=\left( {%
\begin{array}{c}
{{\psi _{2}}(0)} \\
{{\psi _{5}}(0)}%
\end{array}%
}\right) =\left( {%
\begin{array}{c}
{\cos (\theta /2)} \\
{\ -{e^{i\gamma }}\sin (\theta /2)}%
\end{array}%
}\right) ,
\end{equation*}%
we have
\begin{equation*}
\left( {%
\begin{array}{c}
{{\psi _{2}}(t)} \\
{{\psi _{5}}(t)}%
\end{array}%
}\right) ={e^{-iHt}}\left( {%
\begin{array}{c}
{{\psi _{2}}(0)} \\
{{\psi _{5}}(0)}%
\end{array}%
}\right) =\left( {%
\begin{array}{c}
{\cos \frac{\theta }{2}\cos \omega t+i\sin \omega t(\cos \frac{\theta }{2}%
\cos \alpha +{e^{i\gamma }}\sin \frac{\theta }{2}\sin \alpha )} \\
{\ -{e^{i\gamma }}\sin \frac{\theta }{2}\cos \omega t-i\sin \omega t(\cos
\frac{\theta }{2}\sin \alpha -{e^{i\gamma }}\sin \frac{\theta }{2}\cos
\alpha )}%
\end{array}%
}\right) ,
\end{equation*}%
from which we can compute the population or phase difference between two
states at any time. The spin polarization is given by
\begin{equation*}
\langle \sigma _{z}\rangle =\cos \theta +\sin \theta \left( \sin 2\alpha
\sin ^{2}\omega t\cos \gamma -\sin \alpha \sin 2\omega t\sin \gamma \right) ,
\end{equation*}%
which can be written in a sinusoidal form
\begin{equation*}
\langle \sigma _{z}\rangle =\cos \theta +\frac{1}{2}\sin \theta \sin 2\alpha
\cos \gamma -\sin \theta \sin \alpha \sqrt{\sin ^{2}\gamma +\cos ^{2}\alpha
\cos ^{2}\gamma }\sin (2\omega t+\varphi _{0}),
\end{equation*}%
with $\tan \varphi _{0}=\cos \alpha \cot \gamma $. The Rabi oscillation
period is clearly $T=\pi /\omega $, which is independent of initial states
for non-interacting cases within the 2-level approximation.

\begin{figure}[t]
\centering
\includegraphics[width=0.52\textwidth]{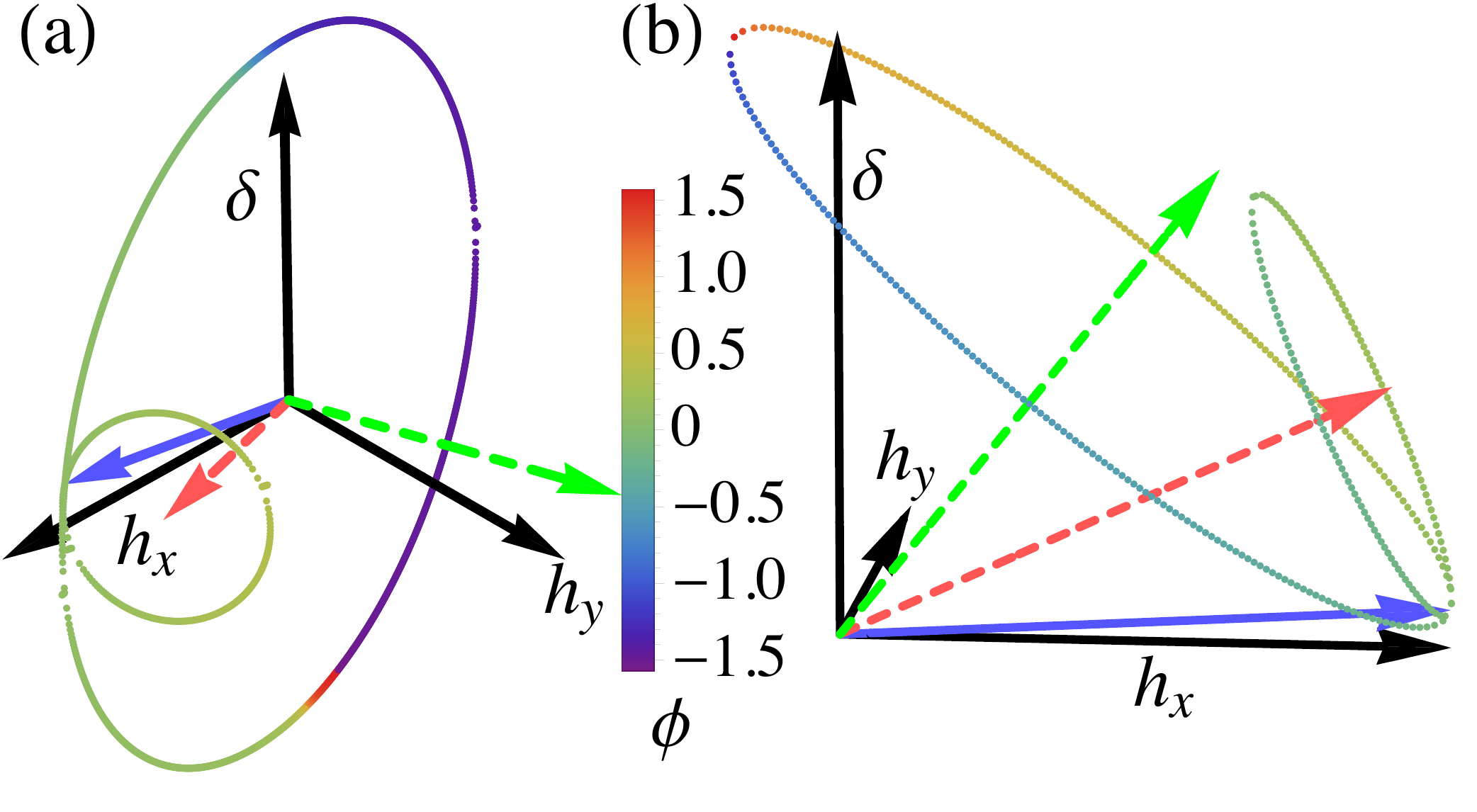}
\caption{(Color online) Cartoon illustration of the quench dynamics when
interaction is weak compared to the coupling strength. Blue solid arrow
denotes the orientation of the initial state, while the two dashed arrows
(red and green) denote the orientation of the effective magnetic fields
after a quench for plasma and Josephson oscillations, respectively. The
circles give the trajectories of moment precession and the color shows the
relative phase $\protect\phi $, which is defined in the $h_{x}$-$h_{y}$
plane. (a) Tunneling-phase-driven MSJJ. The arrows can have arbitrary
orientation in space. (b) Voltage-driven MSJJ. The arrows are confined in a
plane, where their angles to $h_{x}$ remain a constant (for this panel, the
angle is simply $0$).}
\label{figs1}
\end{figure}

It is well known that such Rabi oscillation in a two-level system can be
understood as the precession of a magnetic moment in a magnetic field. For $%
\phi _{L}=0$, the effective magnetic field lies in the $x$-$z$ plane with
strength $\omega $ and angle $\alpha =\arctan \frac{V_{0}}{k_{L}\delta }$
from the $z$-axis. A non-zero $\phi _{L}$ simply gives it a $y$ component.
Both tunneling-phase-driven and voltage-driven processes are illustrated in
Fig.~\ref{figs1}. Since $\phi $ is defined in the $x$-$y$ plane, the system
will undergo a Josephson oscillation when the origin is enclosed by the
projection of the precessing trajectories on the $x$-$y$ plane. A non-zero
detuning is preferred for the oscillation even for the
tunneling-phase-driven process because $\phi $ may otherwise only jump
between $\pm \pi $.

\subsection{Variational analyses}

The atomic interaction can be described in the mean field approximation by
\begin{equation*}
H_{\mathrm{I}}=\frac{1}{2}\int dx\left[ ~g_{\uparrow \uparrow }|\psi
_{\uparrow }|^{4}+g_{\downarrow \downarrow }|\psi _{\downarrow
}|^{4}+2g_{\uparrow \downarrow }|\psi _{\uparrow }|^{2}|\psi _{\downarrow
}|^{2}\right] .
\end{equation*}%
In the following discussions, we assume $g_{\uparrow \uparrow }\sim
g_{\downarrow \downarrow }\sim g$. The spinor wave function considered here
is
\begin{equation*}
\left(
\begin{array}{c}
\psi _{\uparrow } \\
\psi _{\downarrow }%
\end{array}%
\right) =\sqrt{n}\left( C_{2}\left(
\begin{array}{c}
\cos {\theta } \\
-\sin {\theta }%
\end{array}%
\right) e^{ik_{1}x}+C_{5}\left(
\begin{array}{c}
\sin {\theta } \\
-\cos {\theta }%
\end{array}%
\right) e^{-ik_{1}x}\right)
\end{equation*}%
where $2\theta =\arccos \left( {k_{1}/k_{R}}\right) $ can be solved from
minimizing the single particle energy. The energy density given by this
spinor wavefunction is \cite{Li2012}%
\begin{equation*}
\epsilon =\frac{1}{2}k_{R}^{2}-\frac{1}{2}\Omega \sin 2\theta -\Omega
_{L}\cos ^{2}\theta (C_{2}^{\ast }C_{5}e^{i\phi _{L}}+C_{2}C_{5}^{\ast
}e^{-i\phi _{L}})\delta _{k_{1},k_{L}}-\frac{\delta }{2}%
(|C_{2}|^{2}-|C_{5}^{2}|)-F\frac{k_{1}^{2}}{2k_{R}^{2}}+G_{1}\left(
1+2|C_{2}|^{2}|C_{5}|^{2}\right) ,
\end{equation*}%
where $F=(k_{R}-2G_{2})^{2}+4(G_{1}+2G_{2})|C_{2}|^{2}|C_{5}|^{2}$, $%
G_{1}=n(g+g_{\uparrow \downarrow })/4$, and $G_{2}=n(g-g_{\uparrow
\downarrow })/4$.

Through variational methods, one can find $k_{1}=k_{R}\sqrt{1-\Omega
^{2}/(4F)^{2}}$. The equation of motion in the main text can be obtained by
computing the variation of the energy functional with respect to $C_{2}$ and
$C_{5}$, with
\begin{equation*}
g_{G}=-4G_{2}+\frac{1}{2}\Omega ^{2}\frac{(G_{1}+2G_{2})(k_{R}-2G_{2})}{%
\left( \frac{k_{R}-2G_{2}}{|C_{2}|^{2}|C_{5}|^{2}}%
+4|C_{2}|^{2}|C_{5}|^{2}(G_{1}+2G_{2})\right) ^{3}},
\end{equation*}%
which is dependent on the product $|C_{2}|^{2}|C_{5}|^{2}$. However, the
dependency can be eliminated when the interaction strength is weak compared
to $E_{R}$, i.e., $G_{i}\ll k_{R}$, yielding $%
g_{G}=2G_{1}-2(k_{1}^{2}/k_{R}^{2})(G_{1}+2G_{2})$. Taking $C_{j}=\sqrt{N_{j}%
}e^{i\theta _{j}},j=\{2,5\}$ and in terms of the phase difference $\phi
=\theta _{2}-\theta _{5}$ and fractional population difference $z=\frac{%
N_{2}-N_{5}}{N}$, we obtain the classical equations of motions in the main
text
\begin{eqnarray*}
\partial _{t}z &=&-\sqrt{1-z^{2}}\sin \left( \phi -\phi _{Lf}\right) , \\
\partial _{t}\phi &=&\frac{g_{G}}{V_{0}}z+\frac{z}{\sqrt{1-z^{2}}}\cos
\left( \phi -\phi _{Lf}\right) +\frac{k_{L}\delta }{V_{0}},
\end{eqnarray*}%
where time has been rescaled as $t\rightarrow 2V_{0}t$. As long as the
detuning after the quench is large enough, the system is ensured to
experience a Josephson oscillation as $\phi \sim (k_{L}\delta /V_{0})t$. The
classical Hamiltonian can be found through the conjugate relation $\partial
_{t}z=-\frac{\partial H_{c}}{\partial \phi }$ and $\partial _{t}\phi =\frac{%
\partial H_{c}}{\partial z}$,
\begin{equation*}
H_{c}=\frac{g_{G}}{2\tilde{V}}z^{2}-\sqrt{1-z^{2}}\cos \left( \phi -\phi
_{Lf}\right) +\frac{k_{L}\delta }{V_{0}}z+H_{c0},
\end{equation*}%
where $H_{c0}$ is an integration constant.

\subsection{Voltage-driven MSJJ}

\begin{figure}[t]
\centering
\includegraphics[width=0.52\textwidth]{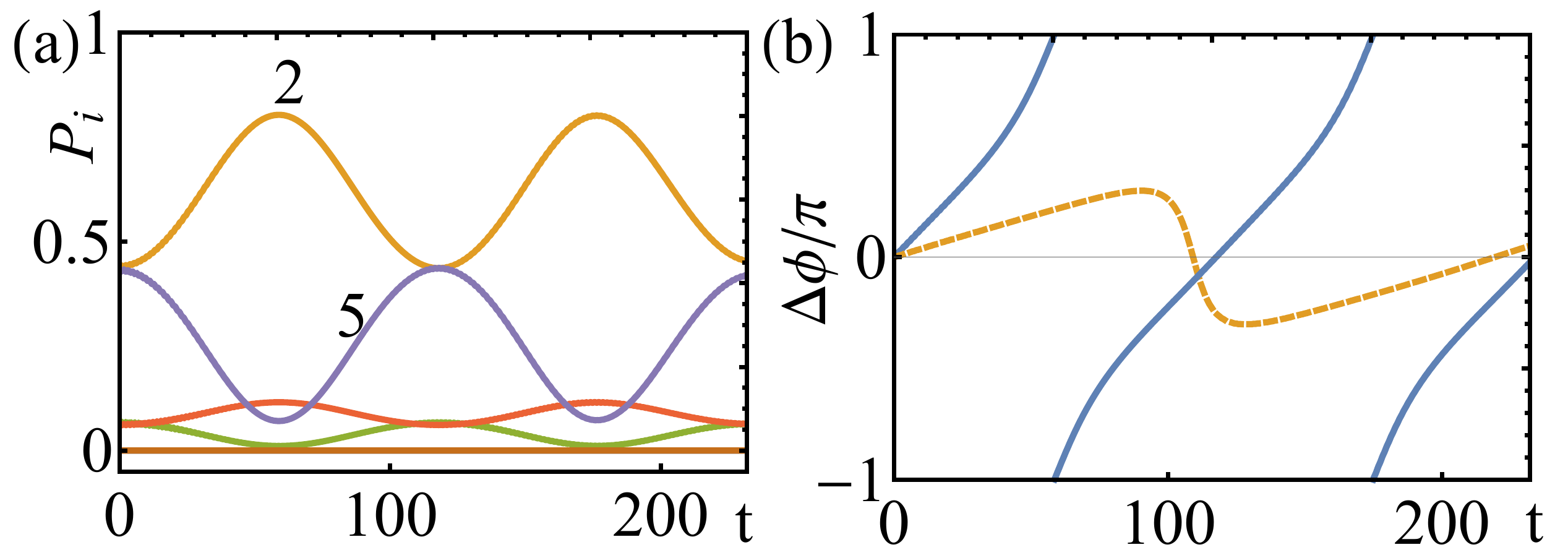}
\caption{(Color online) Similar as Fig.~3 (a) and (b), but for a
voltage-driven MSJJ.}
\label{figs2}
\end{figure}

The GPE simulation results for a voltage(detuning)-driven Josephson
oscillation are shown in Fig.~\ref{figs2}(a), indicating that our $2$-state
approximation still works. The parameters are chosen as $\Omega =2.7$, $%
\Omega _{L}=0.015$, $\phi =0$ and $ng=0.07$, while $\delta $ is quenched
from $0$ Hz to $0.054$. The evolution of its phase difference is plotted in
Fig.~\ref{figs2}(b) as the blue curves. For comparison, we also present the
phase difference for plasma oscillation, where all the parameters are the
same, except that $\delta $ is quenched to $0.016$. Similar self-trapping
effects may also be observed when $\delta $ is quenched from a finite value
to $0$. A constant $\pi $-oscillation occurs when $\phi _{L}=\pi /2$ and $%
\delta $ is quenched from a small number to $0$ (to ensure symmetric
oscillation). In this case, the phase difference is almost a constant at $%
\pi /2$, while the polarization oscillates symmetrically with an observable
amplitude.

\subsection{Numerical methods}

In the numerical simulation, we consider a 1D BEC in a harmonic trap with
spin-orbit coupling and Raman-assisted tunneling. The system is described by
the GPE (1) in the main text. We use imaginary-time evolution to obtain the
ground state numerically with a time-split-operator
numerical method \cite{Bader2013} (see \cite{Bao2003} for alternative methods). The resulting
phase diagram is compared with that generated though variational analysis.
The dynamics are studied through real-time evolution of the ground state
and the time interval $\delta t$ for each evolution step is reduced until
the final state after time $T$ has converged. In our simulation, the
real-space wavefunction is sampled by $2^{11}$ points, and the ground state
energy is achieved to a precision up to $10^{-9}$. We choose $\delta t=0.01$
for each step in the real-time evolution.

\begin{thebibliography}{99}
\bibitem{Josephson1962} B.D. Josephson, {Possible new effects in
superconductive tunnelling}, \href{https://doi.org/10.1016/0031-9163(62)91369-0}%
{Phys. Rev. Lett. \textbf{1}, 251 (1962)}.

\bibitem{Josephson1974} B.D. Josephson, {The discovery of tunnelling
supercurrents}, \href{https://doi.org/10.1103/RevModPhys.46.251}{Rev. Mod.
Phys. \textbf{46}, 251 (1974)}.


\bibitem{Makhlin2001} Y. Makhlin, G. Schon, and A. Shnirman, {Quantum-state
engineering with Josephson-junction devices}, \href{https://doi.org/10.1103/RevModPhys.73.357}%
{Rev. Mod. Phys. \textbf{73}, 357 (2001)}.

\bibitem{Ryu2013} C. Ryu, P. W. Blackburn, A. A. Blinova, and M. G. Boshier,
{Experimental Realization of Josephson Junctions for an Atom SQUID}, \href{https://doi.org/10.1103/PhysRevLett.111.205301}%
{Phys. Rev. Lett. \textbf{111}, 205301 (2013)}.



\bibitem{Martinis2001} J. M. Martinis, S. Nam, J. Aumentado, and C. Urbina, {%
Rabi Oscillations in a Large Josephson-Junction Qubit}, \href{https://doi.org/10.1103/PhysRevLett.89.117901}%
{Phys. Rev. Lett. \textbf{89}, 117901 (2002)}.

\bibitem{Astafiev2006} O. Astafiev, Y. A. Pashkin, Y. Nakamura, T. Yamamoto,
and J. S. Tsai, {Temperature Square Dependence of the Low Frequency $1/f$
Charge Noise in the Josephson Junction Qubits}, \href{https://doi.org/10.1103/PhysRevLett.96.137001}%
{Phys. Rev. Lett. \textbf{96}, 137001 (2006)}.

\bibitem{Martinis2009} J. M. Martinis, M. Ansmann, and J. Aumentado, {Energy
Decay in Superconducting Josephson-Junction Qubits from Nonequilibrium
Quasiparticle Excitations}, \href{https://doi.org/10.1103/PhysRevLett.103.097002}%
{Phys. Rev. Lett. \textbf{103}, 097002 (2009)}.

\bibitem{Paik2011} H. Paik \textit{et al}, {Observation of High Coherence in
Josephson Junction Qubits Measured in a Three-Dimensional Circuit QED
Architecture}, \href{https://doi.org/10.1103/PhysRevLett.107.240501}{Phys.
Rev. Lett. \textbf{107}, 240501 (2011)}. 

\bibitem{Giaever1960} I. Giaever, {Energy Gap in Superconductors Measured by
Electron Tunneling}, \href{https://doi.org/10.1103/PhysRevLett.5.147}{Phys.
Rev. Lett. \textbf{5}, 147 (1960)}.

\bibitem{Likharev1979} K. K. Likharev, {Superconducting weak links}, \href{https://doi.org/10.1103/RevModPhys.51.101}%
{Rev. Mod. Phys. \textbf{51}, 101 (1979)}.


\bibitem{Pereverzev1997} S. V. Pereverzev, A. Loshak, S. Backhaus, J. C.
Davis \& R. E. Packard, {Quantum oscillations between two weakly coupled
reservoirs of superfluid ${}^3$He}, \href{https://www.nature.com/nature/journal/v388/n6641/full/388449a0.html}%
{Nature \textbf{388}, 451 (1997)}.

\bibitem{Hoskinson2005} E. Hoskinson, R. E. Packard \& T. M. Haard, {%
Oscillatory motion: Quantum whistling in superfluid helium-4}, \href{https://doi.org/10.1038/433376a}%
{Nature \textbf{433}, 376 (2005)}.

\bibitem{Wheatley1975} J. C. Wheatley, {Experimental properties of
superfluid ${}^3$He}, \href{https://doi.org/10.1103/RevModPhys.47.415}{Rev.
Mod. Phys. \textbf{47}, 415 (1975)}.

\bibitem{Leggett1975} A. J. Leggett, {A theoretical description of the new
phases of liquid ${}^3$He}, \href{https://doi.org/10.1103/RevModPhys.47.331}{%
Rev. Mod. Phys. \textbf{47}, 331 (1975)}.


\bibitem{Dalfovo1999} F. Dalfovo, S. Giorgini, L. P. Pitaevskii, and
Stringari, S., {Theory of Bose-Einstein condensation in trapped gases},
\href{https://doi.org/10.1103/RevModPhys.71.463}{Rev. Mod. Phys. \textbf{71}%
, 463 (1999)}.

\bibitem{Smerzi1997} A. Smerzi, S. Fantoni, S. Giovanazzi, and S. R. Shenoy,
{Quantum Coherent Atomic Tunneling between Two Trapped Bose-Einstein
Condensates}, \href{https://doi.org/10.1103/PhysRevLett.79.4950}{Phys. Rev.
Lett. \textbf{79}, 4950 (1997)}.

\bibitem{Raghavan1999} S. Raghavan, A. Smerzi, S. Fantoni, and S. R. Shenoy,
{Coherent oscillations between two weakly coupled Bose-Einstein condensates:
Josephson effects, $\pi$ oscillations, and macroscopic quantum self-trapping}%
, \href{https://doi.org/10.1103/PhysRevA.59.620}{Phys. Rev. A \textbf{59},
620 (1999)}.

\bibitem{Williams1999} J. Williams, R. Walser, J. Cooper, E. Cornell, and M.
Holland, {Nonlinear Josephson-type oscillations of a driven, two-component
Bose-Einstein condensate}, \href{https://doi.org/10.1103/PhysRevA.59.R31}{%
Phys. Rev. A \textbf{59}, R31(R) (1999)}.

\bibitem{Ohberg1999} P. Ohberg and S. Stenholm, {Internal Josephson effect
in trapped double condensates}, \href{https://doi.org/10.1103/PhysRevA.59.3890}%
{Phys. Rev. A \textbf{59}, 3890 (1999)}.

\bibitem{Cataliotti2001} F. S. Cataliotti \textit{et al}, {Josephson
Junction Arrays with Bose-Einstein Condensates}, \href{https://doi.org/10.1126/science.1062612}%
{Science \textbf{293}, 843 (2001)}.

\bibitem{Zibold2010} T. Zibold, E. Nicklas, C. Gross, and M. K. Oberthaler, {%
Classical Bifurcation at the Transition from Rabi to Josephson Dynamics},
\href{https://doi.org/10.1103/PhysRevLett.105.204101}{Phys. Rev. Lett.
\textbf{105}, 204101 (2010)}.


\bibitem{Albiez2005} M. Albiez \textit{et al}, {Direct Observation of
Tunneling and Nonlinear Self-Trapping in a Single Bosonic Josephson Junction}%
, \href{https://doi.org/10.1103/PhysRevLett.95.010402}{Phys. Rev. Lett.
\textbf{95}, 010402 (2005)}.

\bibitem{Levy2007} S. Levy, E. Lahoud, I. Shomroni \& J. Steinhauer, {The
a.c. and d.c. Josephson effects in a Bose-Einstein condensate}, \href{https://doi.org/10.1038/nature06186}%
{Nature \textbf{449}, 579 (2007)}.


\bibitem{Valtolina2015} G. Valtolina \textit{et al}, {Josephson effect in
fermionic superfluids across the BEC-BCS crossover}, \href{https://doi.org/10.1126/science.aac9725}%
{Science, 350, 1505 (2015)}.

\bibitem{Burchianti2017} A. Burchianti \textit{et al}, {\ Connecting
dissipation and phase slips in a Josephson junction between fermionic
superfluids}, \href{https://doi.org/10.1103/PhysRevLett.120.025302}{Phys.
Rev. Lett. \textbf{120}, 025302 (2018)}.

\bibitem{Burchianti2} A. Burchianti, C. Fort, and M. Modugno, Josephson
plasma oscillations and the Gross-Pitaevskii equation: Bogoliubov approach
versus two-mode model, \href{https://doi.org/10.1103/PhysRevA.95.023627}{%
Rev. Phys. A \textbf{95}, 023627 (2017)}.

\bibitem{Anderson1967} P. W. Anderson, {Chapter I The Josephson Effect and
Quantum Coherence Measurements in Superconductors and Superfluids}, \href{https://doi.org/10.1016/S0079-6417(08)60119-5}%
{Prog. Low Temp. Phys. \textbf{5}, 5 (1967)}.


\bibitem{Goldman2014} N. Goldman, G. Juzeliunas, P. Ohberg and I. B.
Spielman, {Light-induced gauge fields for ultracold atoms}, \href{https://doi.org/10.1088/0034-4885/77/12/126401}%
{Rep. Prog. Phys. \textbf{77}, 126401 (2014)}.

\bibitem{Lin2011} Y.-J. Lin, K. Jim\'{e}nez-Garc\'{\i}a, and I. B. Spielman,
{Spin-orbit-coupled Bose-Einstein condensates}, \href{http://dx.doi.org/10.1038/nature09887}%
{Nature (London) \textbf{471}, 83 (2011)}.

\bibitem{Zhang2012b} J.-Y. Zhang \textit{et al}, {Collective Dipole
Oscillations of a Spin-Orbit Coupled Bose-Einstein Condensate}, \href{http://dx.doi.org/10.1103/PhysRevLett.109.115301}%
{Phys. Rev. Lett. \textbf{109}, 115301 (2012)}.

\bibitem{Qu2013a} C. Qu, C. Hamner, M. Gong, C. Zhang, and P. Engels, {%
Observation of Zitterbewegung in a spin-orbit-coupled Bose-Einstein
condensate}, \href{http://dx.doi.org/10.1103/PhysRevA.88.021604}{Phys. Rev.
A \textbf{88}, 021604(R) (2013)}.

\bibitem{Olson2014} A. J. Olson \textit{et al}, {Tunable Landau-Zener
transitions in a spin-orbit-coupled Bose-Einstein condensate}, \href{http://dx.doi.org/10.1103/PhysRevA.90.013616}%
{Phys. Rev. A \textbf{90}, 013616 (2014)}.

\bibitem{Hamner2014} C. Hamner \textit{et al}, {Dicke-type phase transition
in a spin-orbit-coupled Bose-Einstein condensate}, \href{http://dx.doi.org/10.1038/ncomms5023}%
{Nat. Commun. \textbf{5}, 4023 (2014)}.

\bibitem{Wang2012} P. Wang \textit{et al}, {Spin-Orbit Coupled Degenerate
Fermi Gases}, \href{http://dx.doi.org/10.1103/PhysRevLett.109.095301}{Phys.
Rev. Lett. \textbf{109}, 095301 (2012)}.

\bibitem{Cheuk2012} L. W. Cheuk \textit{et al}, {Spin-Injection Spectroscopy
of a Spin-Orbit Coupled Fermi Gas}, \href{http://dx.doi.org/10.1103/PhysRevLett.109.095302}%
{Phys. Rev. Lett. \textbf{109}, 095302 (2012)}.

\bibitem{Williams2013} R. A. Williams \textit{et al}, {Raman-Induced
Interactions in a Single-Component Fermi Gas Near an $s$-Wave Feshbach
Resonance}, \href{http://dx.doi.org/10.1103/PhysRevLett.111.095301}{Phys.
Rev. Lett. \textbf{111}, 095301 (2013)}.

\bibitem{Lev} N. Q. Burdick, Y. Tang, and B. L. Lev, {Long-Lived
Spin-Orbit-Coupled Degenerate Dipolar Fermi Gas}, \href{https://doi.org/10.1103/PhysRevX.6.031022}%
{Phys. Rev. X \textbf{6}, 031022 (2016)}.

\bibitem{Jo} B. Song \textit{et al}, {Spin-orbit-coupled two-electron Fermi
gases of ytterbium atoms}, \href{https://doi.org/10.1103/PhysRevA.94.061604}{%
Phys. Rev. A \textbf{94}, 061604(R) (2016)}.

\bibitem{Huang2016} L. Huang \textit{et al}, {Experimental realization of
two-dimensional synthetic spin--orbit coupling in ultracold Fermi gases},
\href{https://doi.org/doi:10.1038/nphys3672}{Nat. Phys. \textbf{12}, 540
(2016)}.

\bibitem{Meng2016} Z. Meng \textit{et al}, {Experimental Observation of a
Topological Band Gap Opening in Ultracold Fermi Gases with Two-Dimensional
Spin-Orbit Coupling}, \href{https://doi.org/10.1103/PhysRevLett.117.235304}{%
Phys. Rev. Lett. \textbf{117}, 235304 (2016)}.

\bibitem{Pan2016} Z. Wu \textit{et al}, {Realization of two-dimensional
spin-orbit coupling for Bose-Einstein condensates}, \href{https://doi.org/10.1126/Science.aaf6689}%
{Science \textbf{354}, 83 (2016)}.


\bibitem{Galitski-review} V. Galitski, and I. B. Spielman, {Spin--orbit
coupling in quantum gases}, \href{https://doi.org/10.1038/nature11841}{%
Nature \textbf{494}, 49 (2013)}.

\bibitem{Zhang2016} Y. Zhang, M. E. Mossman, T. Busch, P. Engels, C. Zhang, {%
Properties of spin--orbit-coupled Bose-Einstein condensates}, \href{https://doi.org/10.1007/s11467-016-0560-y}%
{Front. Phys. \textbf{11}, 118103 (2016)}.


\bibitem{Spielman2015} K. Jim\'{e}nez-Garc\'{\i}a \textit{et al}, {Tunable
Spin-Orbit Coupling via Strong Driving in Ultracold-Atom Systems}, \href{https://doi.org/10.1103/PhysRevLett.114.125301}%
{Phys. Rev. Lett. \textbf{114}, 125301 (2015)}.


\bibitem{BEC-book} C. J. Pethick and H. Smith, Bose-Einstein Condensation in
Dilute Gases, Cambridge University Press (2002).


\bibitem{Li2012} Y. Li, L. P. Pitaevskii, and S. Stringari, {Quantum
Tricriticality and Phase Transitions in Spin-Orbit Coupled Bose-Einstein
Condensates}, \href{https://doi.org/10.1103/PhysRevLett.108.225301}{Phys.
Rev. Lett. \textbf{108}, 225301 (2012)}.


\bibitem{Supp} See Supplementary Materials for derivation of system
Hamiltonian, 2-mode model, voltage-driven MSJJ and numerical recipes for GPE
simulations, which includes Ref. \cite{Bao2003}.

\bibitem{Bao2003} W. Bao, D. Jaksch, P. A. Markowich, {Numerical solution of
the Gross-Pitaevskii equation for Bose-Einstein condensation}, \href{https://doi.org/10.1016/S0021-9991(03)00102-5}%
{J. Comput. Phys. \textbf{187}, 318 (2003)}.


\bibitem{Brion2007} E. Brion, L. H. Pedersen and K. M\o lmer, {Adiabatic
elimination in a lambda system}, \href{https://doi.org/doi:10.1088/1751-8113/40/5/011}%
{J. Phys. A \textbf{40}, 1033 (2007)}.


\bibitem{Bader2013} P. Bader, S. Blanes, and F. Casas, {Solving the Schr\"{o}%
dinger eigenvalue problem by the imaginary time propagation technique using
splitting methods with complex coefficients}, \href{https://doi.org/10.1063/1.4821126}%
{J. Chem. Phys. \textbf{139}, 124117 (2013)}.


\bibitem{Martone2014} G. I. Martone, Y. Li, and S. Stringari, {Approach for
making visible and stable stripes in a spin-orbit-coupled Bose-Einstein
superfluid}, \href{https://doi.org/10.1103/PhysRevA.90.041604}{Phys. Rev. A
\textbf{90}, 041604 (2014)}.

\bibitem{Martone2016} G. I. Martone, T. Ozawa, C. Qu, S. Stringari, {%
Optical-lattice-assisted magnetic phase transition in a spin-orbit-coupled
Bose-Einstein condensate}, \href{https://doi.org/10.1103/PhysRevA.94.043629}{%
Phys. Rev. A \textbf{94}, 043629 (2016)}.

\bibitem{JunRuLi2017} J.-R. Li \textit{et al}, {A stripe phase with
supersolid properties in spin-orbit-coupled {B}ose-{E}instein condensates},
\href{http://doi.org/10.1038/nature21431}{Nature \textbf{543}, 91 (2017)}.


\bibitem{An2017} F. A. An, E. J. Meier, J. Angonga, B. Gadway,{Correlated
dynamics in a synthetic lattice of momentum states}, \href{https://doi.org/10.1103/PhysRevLett.120.040407}%
{Phys. Rev. Lett. \textbf{120}, 040407 (2018).}

\bibitem{Li2016} J. Li \textit{et al}, {Spin-Orbit Coupling and Spin
Textures in Optical Superlattices}, \href{https://doi.org/10.1103/PhysRevLett.117.185301}%
{Phy. Rev. Lett. \textbf{117}, 185301 (2016)}.


\bibitem{Sun2015} K. Sun, C. Qu, and C. Zhang, {%
Spin--orbital-angular-momentum coupling in Bose-Einstein condensates}, \href{http://dx.doi.org/10.1103/PhysRevA.91.063627}%
{Phys. Rev. A \textbf{91}, 063627 (2015)}.

\bibitem{DeMarco2015} M. DeMarco, and H. Pu, {Angular spin-orbit coupling in
cold atoms}, \href{http://dx.doi.org/10.1103/PhysRevA.91.033630}{Phys. Rev.
A \textbf{91}, 033630 (2015)}.

\bibitem{Qu2015} C. Qu, K. Sun, and C. Zhang, {Quantum phases of
Bose-Einstein condensates with synthetic spin--orbital-angular-momentum
coupling}, \href{http://dx.doi.org/10.1103/PhysRevA.91.053630}{Phys. Rev. A
\textbf{91}, 053630 (2015)}.
\end{thebibliography}
\end{document}